\documentclass{article}

\usepackage{arxiv_compact_authors}

\usepackage[utf8]{inputenc} 
\usepackage[T1]{fontenc}    
\usepackage{hyperref}       
\usepackage{url}            
\usepackage{booktabs}       
\usepackage{amsfonts}       
\usepackage{nicefrac}       
\usepackage{microtype}      
\usepackage{lipsum}
\usepackage{graphicx}
\graphicspath{ {./images/} }
\usepackage{amssymb}
\usepackage{amsmath}
\usepackage{multirow}

\usepackage[numbers,sort&compress]{natbib}

\title{Cooperative adsorption and diffusion trapping induced by AlF$_3$ intercalation in graphite}

\author{
H.S. Betancourt-Infante$^{a,*}$,
G.D. Ruano$^{b}$,
F. Bonneto$^{a,c,d}$,
Sindy J. Rodríguez-Sotelo$^{a,*}$
}

\affiliations{

$^{a}$ Instituto de Física del Litoral (IFIS-Litoral, CONICET--UNL), Güemes 3450, 3000 Santa Fe, Argentina

$^{b}$ Dpto.\ de F\'isica de Neutrones, Centro At\'omico Bariloche, Comisi\'on Nacional 
        de Energ\'ia At\'omica (CNEA). Bariloche.

$^{c}$ Departamento de Física, Facultad de Ingeniería Química, Universidad Nacional del Litoral, Santiago del Estero 2829, 3000 Santa Fe, Argentina

$^{d}$ Institute of Environmental Technology (IET), CEET, VSB -- Technical University of Ostrava

$^{1}$ These authors contributed equally.

$^{*}$ Correspondence:
\texttt{heidy.betancourt@santafe-conicet.gov.ar};
\texttt{sindy.rodriguez@santafe-conicet.gov.ar}

}

\begin{document}
\maketitle
\begin{abstract}
Graphite's structural and electronic response to molecular intercalation is central to its performance as a carbon-based electrode material, yet the microscopic coupling between subsurface intercalation and surface adsorption remains poorly understood. We present a first-principles investigation of AlF$_3$ adsorption and intercalation in graphite to explain the microscopic origin of a recently observed two-step self-limiting sorption mechanism. Using density functional theory (DFT-D3), we show that a single intercalated AlF$_3$ molecule locally transforms the structure, electronic properties, and diffusion behavior of graphite through a blister-like surface deformation. Comparing pristine graphite with a graphite surface containing a subsurface intercalated molecule, coverage-dependent adsorption energetics reveal a crossover from repulsive lateral interactions to cooperative binding above the blister, driven by local curvature and intercalation-induced charge redistribution. Diffusion-barrier calculations show that the blister simultaneously acts as a kinetic trap, raising diffusion barriers and transitioning surface mobility from a quasi-barrierless to a thermally activated regime. Charge-density difference and Mulliken population analyses identify the intercalant as a stable electronic reservoir that deepens the surface potential landscape, kinetically immobilizing adsorbed species. Together, these results establish a structure-property relationship for intercalation-induced deformation in graphite, offering a quantitative framework for controlling intercalation efficiency in carbon-based energy storage and conversion systems.
\end{abstract}


\section{Introduction}

Graphite remains a benchmark carbon material for energy storage and conversion applications, not only for conventional Li-ion batteries but also as a versatile platform for studying molecular and ionic intercalation. Its layered structure provides a tunable nanospace where guest species can be accommodated, enabling atomic-scale investigations of how intercalation-induced structural change couples to the electronic properties and interfacial behavior of the host carbon~\cite{dresselhaus2002, wang2021, enoki2003,Yivlianlin,Bouaamlat,Bhauriyal}. Understanding this structure-property coupling is essential for designing electrodes with enhanced stability and capacity, particularly as next-generation battery chemistries demand improved control over surface and subsurface reactions~\cite{zhang2025, whittingham2004, liu2016, he2023,Filoni}.
 
Among emerging intercalants, aluminum-based compounds stand out due to their high theoretical capacity, natural abundance, and low cost~\cite{lin2015, elia2016}. Aluminum trifluoride (AlF$_3$) is especially relevant because of its strong Al--F bonding, molecular stability, and fluorine-rich character, properties already exploited in AlF$_3$ coatings that improve interfacial stability in battery electrodes~\cite{xu2025,li2012}, including graphite anodes that deliver markedly improved cycling retention (92\% vs.\ 81\% after 300 cycles) relative to uncoated graphite~\cite{Ding2012}. Beyond this technological role, AlF$_3$ provides an attractive model system for investigating the fundamental coupling between molecular adsorption, subsurface intercalation, and structural deformation in graphitic materials.
 
Despite this interest, the fundamental relationship between surface adsorption and subsurface intercalation of AlF$_3$ in graphite remains poorly understood. Recent AES and REELS measurements revealed a two-stage sorption mechanism: at low molecular flux, AlF$_3$ gradually intercalates through structural defects, whereas at higher coverage or flux the process transitions to surface adsorption~\cite{Betancourt2026}. These results highlight the importance of substrate crystallinity and defect density in controlling intercalation kinetics and the structural evolution of AlF$_3$ layers.
 
Previous computational studies provided the first atomistic description of AlF$_3$ intercalation, showing that the molecule is thermodynamically stable in stage-1 and stage-2 configurations and induces significant charge transfer to the graphite host~\cite{rodriguez2021}. Molecular dynamics simulations further revealed that intercalated AlF$_3$ can form lateral clusters that generate localized elastic deformations and blister-like structures~\cite{candia2022}. Complementary STM, XPS, and RBS experiments confirmed the presence of such surface deformations~\cite{rodriguez2023}, suggesting that intercalation-induced structural modifications may be more complex than uniform expansion models predict. Beyond AlF$_3$, the diffusion of adsorbed species on graphene and graphite surfaces is itself a well-studied phenomenon with direct implications for thin-film growth and heterostructure fabrication, with reported activation barriers and diffusion regimes strongly dependent on the adsorbate and the local surface environment~\cite{Tang2017,Gervilla2020}.

Building on this framework, the present work investigates the atomistic mechanism by which intercalation-induced graphite deformation modifies subsequent monomeric and dimeric adsorption and surface diffusion. By combining adsorption energetics, diffusion-barrier calculations, and electronic-structure analyses, we identify the microscopic origin of the cooperative adsorption behavior and kinetic trapping associated with blister formation.

These observations point to a more nuanced picture than a simple either/or scenario: rather than treating molecular clustering and progressive intercalation-adsorption coupling as mutually exclusive explanations for the flux-dependent sorption behavior, we test both quantitatively and show that they operate with different relative weight across flux regimes---a distinction developed in detail in Section~\ref{sec:results} once the experimental evidence motivating each hypothesis is presented.
 
Our results reveal that subsurface intercalation induces blister-like deformations that act as cooperative adsorption centers and molecular traps, amplifying charge transfer while suppressing surface diffusion. This blister-induced trapping mechanism provides a microscopic explanation for the biexponential sorption kinetics observed experimentally~\cite{Betancourt2026} and offers a framework for engineering carbon-based electrodes with controlled intercalation efficiency.

\section{Experimental and Computational Methods}

\subsection{Experimental Setup}

The experimental AES and REELS data discussed in this work were reported in Ref.~\cite{Betancourt2026}. Briefly, AlF$_3$ thin films were thermally evaporated under ultra-high vacuum (UHV) conditions onto two HOPG substrates of differing crystalline quality (grades~2 and~3 from SPI), providing lower (LD) and higher (HD) surface densities of defects. AES spectra were acquired using a PHI SAM~590A spectrometer equipped with a cylindrical mirror analyser (CMA) operating in derivative mode, with an energy resolution of 0.3\%. The primary electron beam was set at 3~keV with an incidence angle of $30^{\circ}$ relative to the surface normal. The Al~LMM, F~KLL, and C~KLL Auger transitions were monitored as a function of deposition time, revealing the two-step sorption mechanism that motivates the present computational investigation.

\subsection{Density Functional Theory Calculations}

First-principles calculations were performed using density functional theory (DFT) as implemented in the OpenMX 3.9 code~\cite{ozaki2003}. The exchange-correlation functional was treated within the generalized gradient approximation (GGA) using the Perdew--Burke--Ernzerhof (PBE) parametrization~\cite{perdew1996}. Van der Waals (vdW) interactions were included through the DFT-D3 method of Grimme et al.\ with Becke--Johnson (BJ) damping~\cite{grimme2010, grimme2011}, essential for accurately describing interlayer forces and molecular--surface interactions in graphitic systems~\cite{Bjorkman2012}.

The graphite substrate was modeled using a $7\times6\times1$ Bernal-stacked supercell containing 252 carbon atoms, with lateral dimensions of $17.29\times12.83$~\AA\ and a vacuum region of approximately 25~\AA\ along the $c$-axis to prevent spurious periodic interactions. This supercell model was used throughout all adsorption, intercalation, and diffusion studies.

Norm-conserving pseudopotentials were employed with pseudoatomic orbital (PAO) basis sets specified as Al\,7.0-s2p2d1, C\,6.0-s2p2d1, and F\,6.0-s2p2d1. A real-space energy cutoff of 180~Ry ensured convergence of total energies and atomic forces. For all adsorption and intercalation energy calculations, the Brillouin zone was sampled using a $6\times6\times1$ Monkhorst--Pack $k$-point grid. Self-consistent field (SCF) convergence was achieved with a tolerance of $1\times10^{-7}$~Hartree, and structural optimizations continued until forces on all atoms fell below $1\times10^{-4}$~Hartree/bohr, using the rational function (RF) optimizer.

For the nudged elastic band (NEB) calculations~\cite{henkelman2000}, the $k$-point sampling was reduced to a $4\times4\times1$ grid and the SCF convergence criterion relaxed to $1\times10^{-6}$~Hartree to maintain computational tractability for the large supercell. All atomic degrees of freedom---both the diffusing AlF$_3$ molecule and the graphite substrate---were fully relaxed at every image. The influence of substrate relaxation on the computed barriers was assessed through additional frozen-substrate calculations, in which all carbon atoms were held fixed at their bulk positions; these results are reported in the Supporting Information (Section~S2) for methodological comparison.

\section{Results and Discussion}
\label{sec:results}
\subsection{Dependence of AlF\textsubscript{3} sorption on the 
molecular flux and HOPG defect density}

Understanding the parameters controlling the sorption of AlF$_3$ molecules on graphite is essential for elucidating the interplay between molecular flux, substrate defect density, and the resulting film structure---factors that ultimately govern the performance of AlF$_3$-modified carbon electrodes in battery applications. We tackle this problem by experimentally studying a model system: AlF$_3$ sorption on highly oriented pyrolytic graphite (HOPG) under ultra-high vacuum conditions.
 
\begin{figure*}[!t]
    \centering
    \includegraphics[width=\textwidth]{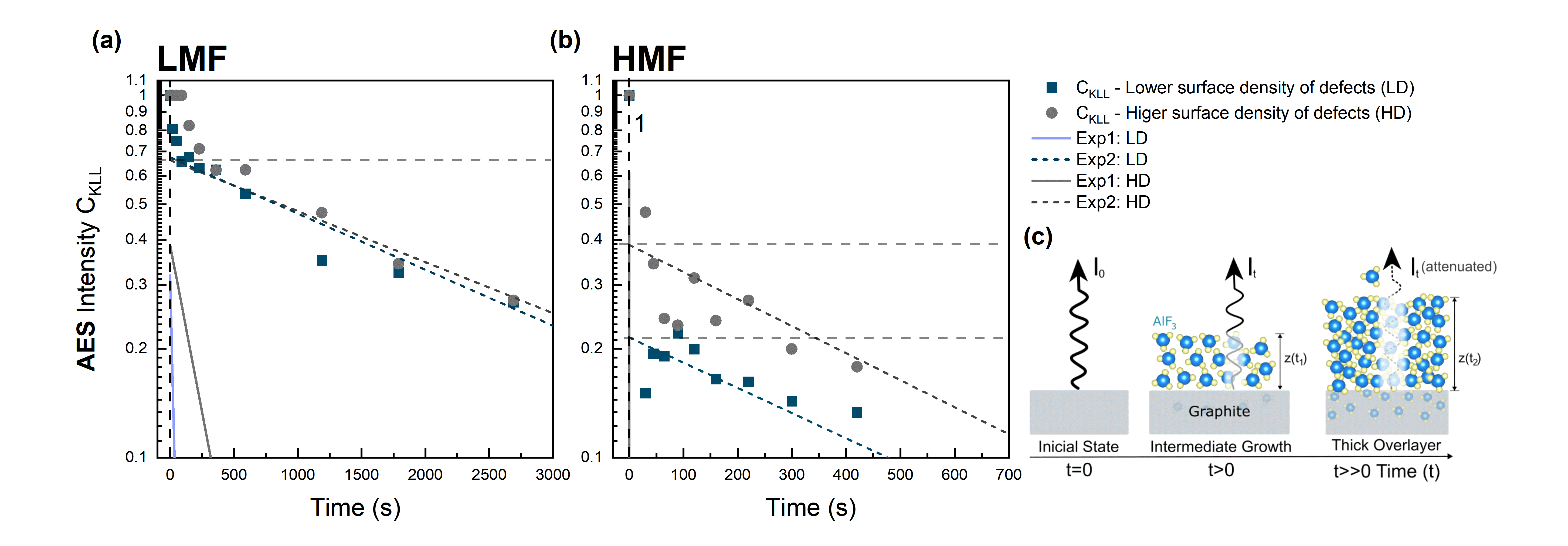}
    \caption{Normalized intensity of C~KLL as a function of deposition time for lower (blue squares, LD) and higher (dark circles, HD) defect density substrates under (a)~low molecular flux (LMF) and (b)~high molecular flux (HMF) regimes. Experimental data (symbols) were fitted using a biexponential decay model (solid lines) with individual exponential components shown in matching colors. (c)~Schematic representation of the evolving overlayer thickness $z(t)$ and the corresponding attenuation of the substrate signal from $I_0$ to $I(t)$.}
    \label{fig_1}
\end{figure*}

AES is a surface-sensitive technique that exploits the short inelastic mean free path (IMFP) of Auger electrons to probe the outermost atomic layers of a material, making it particularly well-suited for studying thin-film growth and interfacial processes~\cite{ruano2011,vidal2015,vaquila2000,pomiro2017}. A particularly useful observable is the normalized substrate signal $I(t)/I_0$, where $I(t)$ reflects the attenuation of the pristine signal $I_0$ by the growing adlayer of thickness $z(t)$ (Fig.~\ref{fig_1}c). Its functional form distinguishes Frank--van der Merwe layer-by-layer growth, which produces steps or kinks in the attenuation curve as observed for C$_{60}$ on Si and Cu~\cite{vidal2015}, from Volmer--Weber island growth or disordered 3D growth, which follows a smooth Beer--Lambert decay as seen for AlF$_3$ on copper~\cite{ruano2011}. This distinction matters directly for interpreting the biexponential behavior below: layer-by-layer growth implies a strongly substrate-mediated process, whereas 3D growth implies weaker lateral coupling and more stochastic site occupation.
 
Thermal evaporation of AlF$_3$ was carried out at two contrasting molecular fluxes, $(5.2\pm0.6)\times10^{12}$ molecules $\cdot$cm$^{-2}\cdot$s$^{-1}$ (low molecular flux, LMF) and $(1.8\pm0.2)\times10^{13}$ molecules $\cdot$cm$^{-2}\cdot$s$^{-1}$ (high molecular flux, HMF), onto two HOPG substrates with lower (LD, grade~2 SPI) and higher (HD, grade~3 SPI) surface densities of defects. Full experimental details are reported elsewhere~\cite{Betancourt2026}.

Figure~\ref{fig_1} shows the C~KLL attenuation in semilog scale for both flux regimes and substrates. Simple exponential behavior, expected from Beer--Lambert attenuation in uniform layer-by-layer growth, is immediately ruled out: all curves are instead well described by a biexponential model, indicating the coexistence---rather than sequential replacement---of two distinct sorption mechanisms operating on different timescales throughout deposition.
 
Under LMF conditions (Fig.~\ref{fig_1}a), the first exponential component (Exp1, lighter shades) is strongly substrate-dependent and is associated with intercalation through step edges, grain boundaries, and structural defects that act as entry channels~\cite{candia2022,rodriguez2023}: since defect density differs substantially between HD and LD substrates, a mechanism gated by defect availability should---and does---show pronounced substrate dependence. The second component (Exp2, darker shades) is substrate-independent and reflects Beer--Lambert-like accumulation of an AlF$_3$ overlayer once interlayer sites accessible at that flux have saturated---the growth mode expected once defect-mediated intercalation is no longer dominant.

This coexistence, rather than a clean sequential handoff, is itself physically significant, and is reinforced by a simple capacity argument: considering a conservative sticking factor of 1 and a flux on the order of $10^{12}$~molecules$\cdot$cm$^{-2}\cdot$s$^{-1}$, with a surface atomic density of roughly $10^{15}$~atom$\cdot$cm$^{-2}$ for graphite, a coverage of 0.1~monolayer of AlF$_3$ is reached after only $\sim$100~s on the outer surface. STM measurements on HOPG substrates of comparable defect density report step-edge densities of order $10^1$--$10^2$~step-edges/cm$^2$~\cite{candia2022}, each providing access to one or more additional interlayer galleries beyond this outer surface. If the same incident flux is shared among all of these galleries, each receives a proportionally smaller effective flux, so that filling every gallery to 0.1~coverage takes proportionally longer than filling the outer surface alone: from $\sim$1~hour, assuming a single gallery per step edge, up to $\sim$1~day if each step edge instead provides access to several stacked galleries---one to three orders of magnitude longer than the observed fast-component timescale (tens to hundreds of seconds).

This large discrepancy indicates that intercalation is kinetically limited well before thermodynamic saturation of the interlayer volume~\cite{rodriguez2021,he2023}, ruling out simple pore-filling as the limiting factor and providing the first experimental signature of the adsorption--intercalation coupling motivating hypothesis~(b) below which is discussed later in this section.
 
Under HMF conditions (Fig.~\ref{fig_1}b), both components change substantially in relative weight rather than either disappearing: the fast component becomes dominant, with preexponential factors nearly double those of LMF (LD:~0.79 vs.~0.32; HD:~0.68 vs.~0.39) and time constants several orders of magnitude larger (see Table 2 in Ref.~\cite{Betancourt2026}), while the slow component shows faster coverage of the LD surface relative to HD---consistent with LD, having fewer intercalation channels, transitioning more rapidly into the adsorption-dominated regime.  This flux dependence is further underscored by comparing equivalent doses: at 1000~s, LMF attenuates the signal to $\sim$50\% for both substrates, whereas the equivalent HMF dose (285~s) reduces it to only 15\% and 25\%, respectively---confirming that the growth kinetics depend on the rate of molecular delivery, not merely on the total dose accumulated.

Taken together, these results show that molecule--molecule and molecule--substrate interactions, coupled to the competition between impinging flux and diffusion rate, operate continuously across both regimes rather than switching between disconnected mechanisms. This motivates two hypotheses, tested quantitatively below rather than treated as mutually exclusive: (a)~enhanced surface coverage under HMF is partly driven by stabilization of molecular clusters relative to isolated monomers; and (b)~intercalation occurs under both flux conditions but progressively conditions the adsorption landscape---more markedly under HMF, to a lesser but measurable extent under LMF---quenching further interlayer penetration as the adsorbed population grows. To test the first, we examine pristine graphite (basal-plane adsorption only); to test the second, we model a graphite surface containing a subsurface intercalated AlF$_3$ molecule that generates a blister-like deformation.


\subsection{Coverage-Dependent Adsorption on Pristine Graphite}
\label{sec:HMF}
To establish the adsorption behavior of AlF$_3$ in the absence of intercalation-induced surface modifications, and thereby provide the reference system required to test hypothesis~(a), adsorption on pristine graphite was systematically investigated for coverages ranging from 1 to 6 molecules. Aluminum atoms were placed above surface carbon atoms (top site), the adsorption geometry previously identified as most stable for AlF$_3$ on graphite~\cite{rodriguez2021}, and fully relaxed from various starting heights. The adsorption energy per molecule is defined as:
 
\begin{equation}
E_{\mathrm{ads}} = \frac{E_{\mathrm{AlF_3+graphite}} - E_{\mathrm{graphite}} - n \cdot E_{\mathrm{AlF_3}}}{n}
\label{eq:eads}
\end{equation}
 
where $E_{\mathrm{AlF_3+graphite}}$ is the total energy of the system with $n$ adsorbed molecules, $E_{\mathrm{graphite}}$ is the energy of the pristine substrate, and $E_{\mathrm{AlF_3}}$ is the energy of an isolated AlF$_3$ molecule.
 
\paragraph{Monomeric Adsorption}

\begin{figure*}[htpb]
    \centering
    \includegraphics[scale=.40]{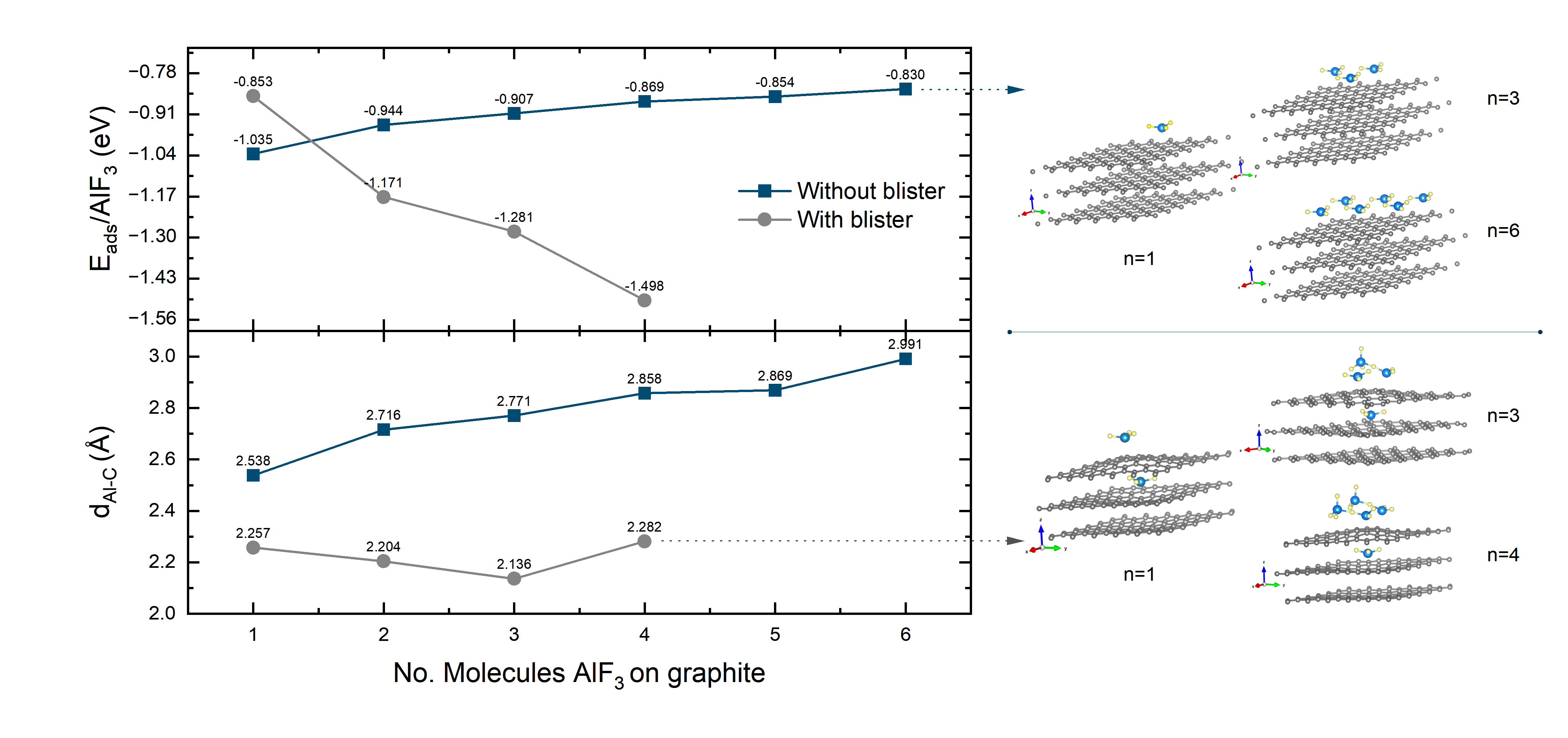}
    \caption{Crossover from repulsive to cooperative adsorption. Top panel: evolution of adsorption energy per AlF$_3$ molecule on pristine graphite (blue squares) and blistered graphite (grey circles). While the pristine surface exhibits a monotonic weakening due to lateral dipole--dipole repulsion, the blistered surface shows a cooperative strengthening with increasing coverage. Bottom panel: corresponding equilibrium Al--C distances. The systematically shorter distances on the blistered surface (2.14--2.28~\AA\ vs.\ 2.54--2.99~\AA\ on pristine graphite) provide direct structural evidence of the stronger surface interaction driven by local curvature and subsurface deformation.}
    \label{fig:energetics_crossover}
\end{figure*}

The results reveal a systematic dependence of adsorption energetics on molecular coverage, consistent with lateral interaction effects typical of physisorbed molecular overlayers~\cite{campbell2012}. Single-molecule adsorption yields a binding energy of $-1.035$~eV with an optimized Al--C distance of 2.538~\AA, comparable in magnitude to adatom binding energies reported for other metal species on graphitic surfaces~\cite{Chan2008_adsorption}. As coverage increases, the adsorption energy per molecule decreases monotonically---from $-1.035$~eV at $n=1$ to $-0.907$~eV at $n=3$, stabilizing in the range $-0.869$ to $-0.830$~eV for $n=4$--6---accompanied by a systematic increase in the average Al--C distance from 2.538~\AA\ to 2.991~\AA\ (Figure~\ref{fig:energetics_crossover}, bottom panel).
 
This monotonic weakening is consistent with the accumulation of repulsive lateral interactions among co-adsorbed molecules: Mulliken analysis (Section~\ref{sec:CDD}) shows that each AlF$_3$ molecule acquires a net negative charge from the graphite surface, establishing a molecule--substrate dipole whose lateral electrostatic repulsion is a well-documented driver of coverage-dependent weakening and increased equilibrium adsorption height in DFT studies of polar molecular adlayers~\cite{Deshlahra2012}. The stabilization of the binding energy beyond $n=4$ marks the onset of a quasi-saturated monolayer.
 
\paragraph{Dimeric Species}
\label{sec:dimers_pristine}
 
To test whether molecular clustering during the gas-phase deposition stage could account for the enhanced coverage rates observed under HMF---hypothesis~(a)---we investigated the spontaneous formation and adsorption of AlF$_3$ dimers on the pristine graphite surface. Two configurations were examined: a single dimer (2~molecules, 1~dimeric unit) and two dimers (4~molecules, 2~dimeric units). The adsorption energetics are presented in Table~\ref{tab:dimers_pristine}.
 
\begin{table}[ht]
\centering
\caption{Adsorption energetics for AlF$_3$ dimers on pristine graphite surfaces.}
\label{tab:dimers_pristine}
\resizebox{0.7\columnwidth}{!}{%
\begin{tabular}{cccc}
\hline
Configuration & Al--C dist. (\AA) & $E_{\mathrm{ads}}$/mol (eV) & $E_{\mathrm{ads}}$/dimer (eV) \\
\hline
1 dimer (2 mol)  & 2.258 & $-0.814$ & $-1.628$ \\
2 dimers (4 mol) & 2.341 & $-0.695$ & $-1.391$ \\
\hline
\end{tabular}}
\end{table}
 
The dimeric species exhibit binding energies per molecule of $-0.814$~eV (1~dimer) and $-0.695$~eV (2~dimers)---comparable to, but not stronger than, isolated monomers at equivalent coverages ($-0.944$~eV for $n=2$, Table~\ref{tab:adsorption_comparison}). The shorter Al--C distances for dimers (2.258--2.341~\AA) suggest that the internal Al--F--Al bridge bonding redistributes electron density in a way that partially compensates for the loss of individual molecule--substrate contact. Importantly, these dimeric configurations did not form spontaneously on the pristine surface: starting from two independent monomers in proximity, structural relaxation kept them as separate adsorbates, and the dimer geometry required manual construction as an initial condition.
 
Taken together, these results show that molecular clustering does not by itself stabilize adsorption on pristine graphite, even in its most favourable geometric configuration: neither spontaneous monomer accumulation nor preformed dimers offer any energetic advantage over isolated monomers. Hypothesis~(a) is therefore not supported as an independent, thermodynamically driven pathway at the level of the pairwise Al--C and Al--F--Al interactions probed here. This does not exclude a purely kinetic role for transient clustering during real deposition---but it does mean that whatever drives the enhanced coverage rate under HMF must originate elsewhere: intercalation itself, as observed experimentally, still occurs under HMF, only to a lesser extent. We now test hypothesis~(b): whether this blister transforms the surface in a way that favours the accumulation of subsequent AlF$_3$ molecules.

\subsection{Adsorption on Intercalation-Induced Blistered Graphite}
\label{sec:LMF}

To test hypothesis~(b), namely that subsurface intercalation progressively conditions the surface adsorption landscape, we considered a graphite surface containing a subsurface intercalated AlF$_3$ molecule. 
This model represents the intercalation-conditioned sorption regime, whose relative contribution becomes increasingly important as the molecular flux decreases toward LMF conditions. 

\begin{figure*}[!t]
    \centering
\includegraphics[width=0.9\columnwidth]{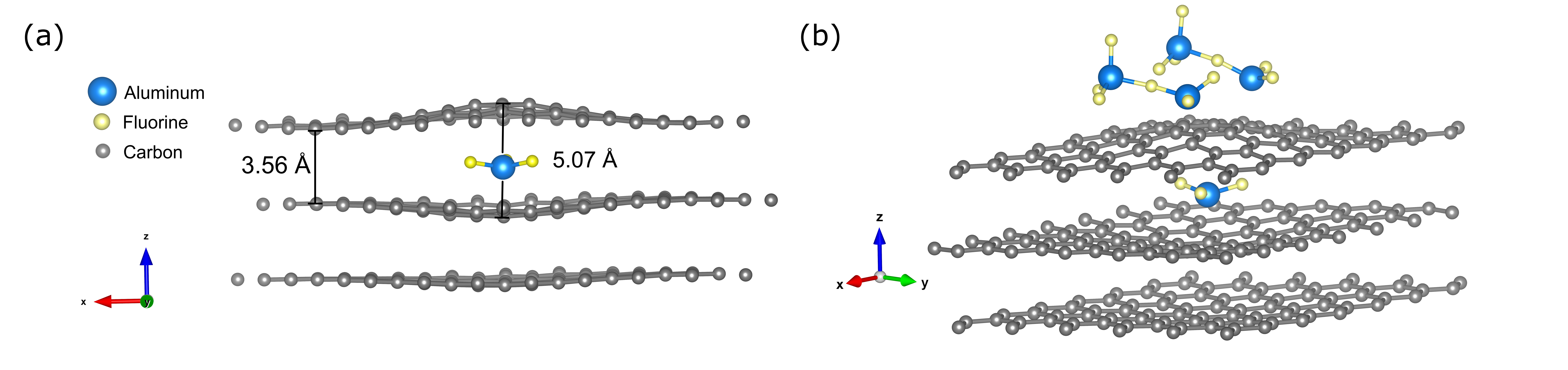}
     \caption{Intercalation-induced blister formation and saturation in graphite. (a)~Side view of the optimized AlF$_3$-intercalated structure. The intercalant induces a localized out-of-plane deformation of the graphene layers, leading to a non-uniform expansion of the interlayer spacing. No C--C bond breaking is observed, indicating elastic accommodation of the intercalant within the graphite lattice. (b)~Fully occupied blister at $n=4$, showing the two Al$_2$F$_6$ dimers occupying the available surface area above the intercalated molecule.}
    \label{fig:blister}
\end{figure*}

The intercalation of AlF$_3$ into Bernal-stacked graphite was systematically explored by varying the interlayer separation; the energetically most favourable configuration was obtained for an initial interlayer separation of 6~\AA. Upon relaxation, the intercalant induces a localized blister-like deformation of the graphene layers, consistent with previous theoretical and experimental studies~\cite{candia2022,rodriguez2023,Bouaamlat}. The resulting structure is characterized by three features: (i)~\textit{localized expansion}---the interlayer distance directly above the intercalant increases from the pristine value of 3.35~\AA\ to 5.07~\AA; (ii)~\textit{gradient decay}---the spacing recovers gradually toward the supercell edges, reaching $\sim$3.56~\AA; and (iii)~\textit{preserved structural integrity}---no C--C bond breaking is observed, indicating purely elastic accommodation of the intercalant within the graphite lattice (Figure~\ref{fig:blister}). 

This deformation creates a unique environment where subsurface and surface interactions coexist. In sharp contrast to pristine graphite, two AlF$_3$ monomers placed on the blistered surface spontaneously dimerize into an Al$_2$F$_6$ unit upon structural relaxation---a behavior not observed on the pristine surface, where dimeric configurations required manual construction  (Section~\ref{sec:dimers_pristine}). The resulting dimer exhibits a characteristic asymmetric geometry: one Al atom remains in close proximity to the surface (2.20~\AA) while the bridging Al atom sits at a larger distance (3.63~\AA). Subsequent addition of molecules follows the sequence $n=1$~(monomer) $\rightarrow$ $n=2$~(dimer) $\rightarrow$ $n=3$~(dimer\,+\,monomer) $\rightarrow$ $n=4$ (2~dimers), at which point the blister region is fully occupied (Figure~\ref{fig:blister}b). Spontaneous dimerization and saturation at $n=4$ are themselves signatures of the enhanced, spatially confined binding environment created by the blister.

To evaluate the cooperative effect of the blister on surface adsorption, AlF$_3$ molecules were sequentially added and fully relaxed, with adsorption energies computed via Equation~\eqref{eq:eads}. For $n=1$, the Al--C distance is 2.257~\AA\ and the adsorption energy is $-0.853$~eV---slightly weaker than on the pristine surface ($-1.035$~eV)---indicating that the blister alone does not immediately enhance binding. However, a clear cooperative effect emerges with increasing coverage: at $n=2$ the energy strengthens to $-1.171$~eV (Al--C: 2.204~\AA), already exceeding the pristine value ($-0.944$~eV); at $n=3$ it reaches $-1.281$~eV (Al--C: 2.136~\AA); and at $n=4$ the system attains its strongest binding of $-1.498$~eV (Al--C: 2.282~\AA), with the slight distance increase at full coverage consistent with incipient steric repulsion between the two dimeric units. This progressive strengthening contrasts sharply with the monotonic weakening observed on pristine graphite (Figure~\ref{fig:energetics_crossover}, Table~\ref{tab:adsorption_comparison}).

\begin{table}[h]
\centering
\caption{Representative adsorption geometries and adsorption energies for systems containing $n=1$ and $n=2$ adsorbed AlF$_3$ molecules on pristine (HMF) and blistered (LMF) graphite. Results for larger numbers of adsorbed AlF$_3$ molecules are presented in Figure~\ref{fig:energetics_crossover}.}
\label{tab:adsorption_comparison}
\setlength{\tabcolsep}{4pt}
\renewcommand{\arraystretch}{1.05}
\resizebox{0.7\columnwidth}{!}{%
\begin{tabular}{cccc}
\toprule
No. molecules  & Surface model & Al--C distance (\AA) & $E_{\mathrm{ads}}$ (eV) \\
\midrule
\multirow{2}{*}{$1$}
& Pristine (HMF)  & 2.538 & $-1.035$ \\
& Blistered (LMF) & 2.257 & $-0.853$ \\
\midrule
\multirow{2}{*}{$2$}
& Pristine (HMF)  & 2.716 & $-0.944$ \\
& Blistered (LMF) & 2.204 & $-1.171$ \\
\bottomrule
\end{tabular}}
\end{table}

These results establish that blisters act as coverage-activated cooperative adsorption centers, driving a transition from a repulsion-dominated regime on pristine graphite to a curvature-assisted cooperative regime on blistered surfaces. This crossover provides the first computational evidence in support of hypothesis~(b): intercalation does not merely precede surface adsorption but actively conditions it through structural modification of the graphite surface, consistent with the coupling between flux and intercalation-adsorption behavior identified experimentally. The energetic picture alone, however, only establishes where AlF$_3$ molecules preferentially adsorb---it does not yet address whether they remain kinetically confined once adsorbed, a distinction essential for connecting this structural model to the AES saturation behavior. We examine this kinetic dimension next through NEB calculations.

\subsection{Diffusion Pathways and Energy Barriers}
\label{sec:diffusion}

\begin{figure*}[!t]
\centering
\includegraphics[width=0.85\textwidth]{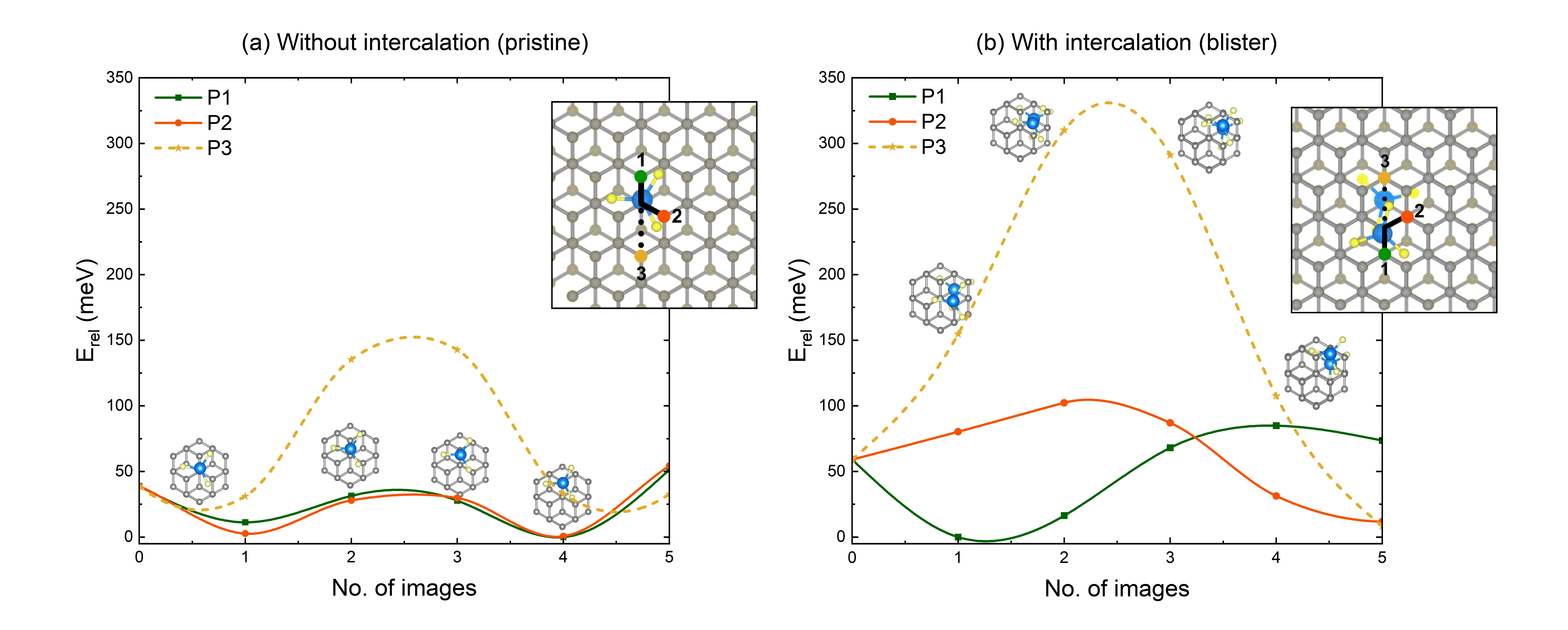}
\caption{NEB energy profiles for AlF$_3$ surface diffusion on (a)~pristine graphite and (b)~blistered graphite, computed with a fully relaxed substrate. Each panel shows $E_{\mathrm{rel}}$ as a function of NEB image index for paths P1--P3. Energy profiles are shifted so that the global minimum coincides with zero; both panels share the same energy scale. Insets show the diffusion pathways on the hexagonal graphite lattice. Representative geometries along P1 (panel~a) illustrate the near-barrierless C--C bond crossing on the pristine surface; geometries along P3 (panel~b) illustrate the strongly activated hexagon-crossing jump above the intercalated blister.}
\label{fig:NEB_profiles_free}
\end{figure*}

Having established the energetic component of hypothesis~(b), namely the cooperative strengthening of adsorption above the blister, we now test its kinetic component by examining whether the subsurface intercalant also suppresses the lateral diffusion of surface AlF$_3$ molecules. NEB calculations were performed along three crystallographically distinct diffusion pathways (P1--P3, Figure~\ref{fig:NEB_profiles_free}) on both surface models, with jump rates estimated via the Eyring--Kramers expression:
 
\begin{equation}
  k = \nu\,\exp\!\left(-\frac{E^{\ddagger}}{k_{\mathrm{B}}T}\right),
  \label{eq:rate}
\end{equation}
 
where $\nu = 10^{12}$~Hz~\cite{campbell2012} is the attempt frequency typical for physisorbed molecules on graphite, $E^{\ddagger}$ is the activation barrier, $k_{\mathrm{B}}$ is the Boltzmann constant, and $T = 300$~K. The effective 2D diffusion coefficient was computed as:
 
\begin{equation}
  D = \frac{1}{4}\sum_{i} d_i^{2}\,k_i,
  \label{eq:diff2d}
\end{equation}

where $d_i$ is the real-space displacement of the Al atom along pathway $i$, extracted from the Cartesian coordinates of the relaxed NEB images. This expression follows the standard random-walk relation between diffusivity and hopping rate~\cite{AlaNissila2002}, with the sum running over the three pathways P1--P3 (Figure~\ref{fig:NEB_profiles_free}, inset) and using the forward rate constants $k_i$. Full results are compiled in Table~\ref{tab:neb_sb_free}.
 
\begin{table*}[ht]
\centering
\caption{NEB diffusion parameters for AlF$_3$ surface migration on pristine and blistered graphite (fully relaxed substrate). $d_{\mathrm{Al}}$: real-space displacement of the Al atom; $E^{\ddagger}$: forward activation barrier; $\Delta E$: reaction energy (product relative to precursor); $k(300\,\mathrm{K})$: jump rate; $\tau = 1/k$: hop time; $k_{\mathrm{B}}T = 25.85$~meV at 300~K.}
\label{tab:neb_sb_free}
\begin{tabular}{cccccccc}
\hline
Surface Model & Path & $d_{\mathrm{Al}}$ (\AA) & $E^{\ddagger}$ (meV) & $\Delta E$ (meV) & $k(300\,\mathrm{K})$ (s$^{-1}$) & $\tau$ & $E^{\ddagger}/k_{\mathrm{B}}T$ \\
\hline
\multirow{3}{*}{Pristine}
& P1 & 1.39 & 12.4  & $+$12.4 & $6.2\times10^{11}$ & 1.6~ps   & 0.48 \\
& P2 & 1.48 & 15.5  & $+$15.5 & $5.5\times10^{11}$ & 1.8~ps   & 0.60 \\
& P3 & 3.07 & 103.9 & $-$6.2  & $1.8\times10^{10}$ & 55.8~ps  & 4.02 \\
\hline
\multirow{3}{*}{Blister}
& P1 & 1.26 & 25.8  & $+$14.5 & $3.7\times10^{11}$ & 2.7~ps      & 1.00 \\
& P2 & 1.66 & 43.2  & $-$47.3 & $1.9\times10^{11}$ & 5.3~ps      & 1.67 \\
& P3 & 3.15 & 251.0 & $-$50.8 & $6.1\times10^{7}$  & 16.5~ns & 9.71 \\
\hline
\multicolumn{2}{l}{$D_{\mathrm{pristine}}(300\,\mathrm{K})$}
& \multicolumn{6}{l}{$\approx 6.4\times10^{-9}$~m$^2$\,s$^{-1}$} \\
\multicolumn{2}{l}{$D_{\mathrm{blister}}(300\,\mathrm{K})$}
& \multicolumn{6}{l}{$\approx 2.8\times10^{-9}$~m$^2$\,s$^{-1}$} \\
\hline
\end{tabular}
\end{table*}
 
On pristine graphite (Figure~\ref{fig:NEB_profiles_free}a), paths P1 and P2 exhibit barriers of 12.4 and 15.5~meV, both below $k_{\mathrm{B}}T = 25.85$~meV at room temperature, with hop times of 1.6 and 1.8~ps indicating essentially barrierless thermal diffusion. The precursor P0 occupies an $\alpha$ hollow site (centred above a carbon of the lower graphene layer); P1 and P2 correspond to nearest-neighbour C--C bond crossings, while P3 constitutes a second-nearest-neighbour jump across the hexagon centre~\cite{AlGraphene2023}, with a barrier of 103.9~meV ($E^{\ddagger}/k_{\mathrm{B}}T = 4.02$) and a hop time of 55.8~ps. The resulting diffusion coefficient $D_{\mathrm{pristine}} \approx 6.4\times10^{-9}$~m$^2$\,s$^{-1}$ is characteristic of a highly mobile physisorbed species.
 
The presence of the subsurface intercalated AlF$_3$ molecule induces a qualitative change in the diffusion landscape (Figure~\ref{fig:NEB_profiles_free}b). The precursor now occupies a $\beta$ site (centred above the interlayer void), reflecting the broken symmetry introduced by the intercalant; the slight inequivalence between $\alpha$ and $\beta$ sites originates from the ABAB stacking symmetry of graphite~\cite{dresselhaus2002}. All barriers increase substantially: P1 and P2 rise to 25.8 and 43.2~meV ($E^{\ddagger}/k_{\mathrm{B}}T = 1.0$ and 1.7), remaining thermally accessible, while P3 reaches 251.0~meV with a hop time of $\sim$16.5~ns, making this pathway strongly suppressed relative to pristine graphite at room temperature.

\begin{figure*}[ht]
\centering
\includegraphics[width=0.85\textwidth]{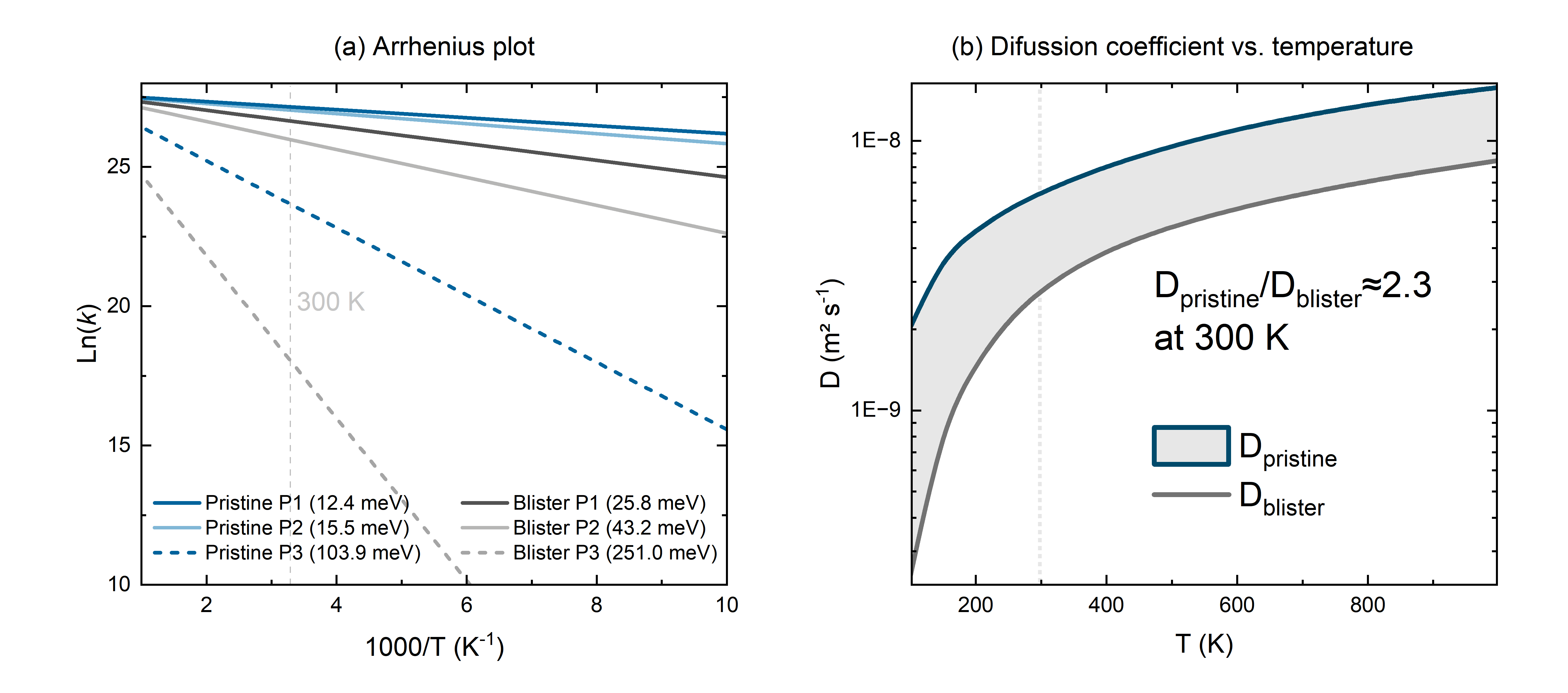}
\caption{Kinetic analysis of AlF$_3$ surface diffusion with fully relaxed substrate. (a)~Arrhenius plot ($\ln k$ vs.\ $1000/T$) for pristine graphite (blue, $E^{\ddagger}_{\mathrm{min}} = 12.4$~meV, $E^{\ddagger}_{\mathrm{max}} = 103.9$~meV) and blistered graphite (gray, $E^{\ddagger}_{\mathrm{min}} = 25.8$~meV, $E^{\ddagger}_{\mathrm{max}} = 251.0$~meV). Solid lines indicate the most favourable diffusion path; dashed lines the least favourable. The vertical dotted line marks $T = 300$~K. (b)~Effective 2D diffusion coefficient $D$ as a function of temperature for pristine (blue) and blistered (gray) graphite, computed via Eq.~(\ref{eq:diff2d}). The shaded area highlights the suppression of surface mobility. At 300~K, $D_{\mathrm{pristine}}/D_{\mathrm{blister}} \approx 2.3$; the ratio increases to $\sim$8 at 100~K.}
\label{fig:arrhenius_D_free}
\end{figure*}

Notably, the negative reaction energies for P2 and P3 ($\Delta E = -47.3$ and $-50.8$~meV, respectively) indicate that the product sites reached along these pathways are more stable than the precursor, implying reverse barriers substantially larger than the forward ones: $E^{\ddagger}_{\mathrm{back}} = E^{\ddagger} - \Delta E$ amounts to 90.5~meV for P2 and 301.8~meV ($\sim$11.7~$k_{\mathrm{B}}T$) for P3---effectively irreversible at room temperature. Once an AlF$_3$ molecule reaches these more stable sites, the thermodynamic driving force thus translates into a kinetic barrier against return, reinforcing the confinement of surface molecules above the intercalated blister.
The resulting $D_{\mathrm{blister}} \approx 2.8\times10^{-9}$~m$^2$\,s$^{-1}$ corresponds to a 2.3-fold reduction in surface mobility relative to pristine graphite. This contrast is directly visible in the Arrhenius representation (Figure~\ref{fig:arrhenius_D_free}a): the steeper slopes of the blister pathways, reflecting their larger activation barriers, cause the pristine and blister trend lines to diverge increasingly as temperature decreases, since the exponential dependence of the jump rate on $E^{\ddagger}/k_{\mathrm{B}}T$ amplifies barrier differences at low $T$. This divergence is quantified directly in terms of mobility in Figure~\ref{fig:arrhenius_D_free}b, where the effective diffusion coefficient $D$ rises with temperature for both systems, as expected from the underlying Arrhenius behaviour, while the shaded region marks the separation between $D_{\mathrm{pristine}}$ and $D_{\mathrm{blister}}$: this gap is narrow near room temperature but widens substantially at low $T$, reflecting a diffusivity ratio that grows from 2.3-fold at 300~K to $\sim$8-fold at 100~K. Frozen-substrate calculations yield a ratio of $\sim$52 at 300~K (Supporting Information, Section~S2), confirming the qualitative conclusion while illustrating the sensitivity to the substrate approximation.
 
The increase in barriers arises from three cooperative effects induced by the subsurface intercalant: (i)~charge redistribution at the graphite surface, which deepens the electrostatic potential wells at hollow sites directly above the intercalant, as detailed in Section~\ref{sec:CDD}; (ii)~mechanical strain in the curved graphene layers~\cite{pereira2009,khestanova2016,Bouaamlat}, which stabilizes the blister apex as a local energy minimum; and (iii)~enhanced Al--C bonding evidenced by shorter equilibrium distances on the blistered surface (2.14--2.28~\AA\ vs.\ 2.54--2.99~\AA\ on pristine graphite, Table~\ref{tab:adsorption_comparison}), translating into a deeper adsorption well and a higher escape barrier.
 
This kinetic landscape provides a direct microscopic rationalization of the biexponential AES dynamics (Figure~\ref{fig_1}). Under LMF conditions, incoming AlF$_3$ molecules diffuse rapidly across the pristine graphite surface with quasi-barrierless mobility ($D \approx 6.4\times10^{-9}$~m$^2$\,s$^{-1}$, hop times of 1--2~ps), enabling efficient exploration and intercalation through structural defects---consistent with the fast, substrate-dependent first exponential component observed experimentally. As intercalation proceeds and blister structures proliferate, barriers rise from $\lesssim k_{\mathrm{B}}T$ to 1--10~$k_{\mathrm{B}}T$, reducing surface mobility by a factor of $\sim$2.3 and trapping molecules above intercalated regions. This progressive kinetic suppression forces subsequent molecular arrivals to accumulate as surface adsorbates rather than continuing to intercalate, explaining the crossover to the slower, substrate-independent second exponential component in the biexponential AES fits (Figure~\ref{fig_1}, panels~a--b).
 
Taken together, the adsorption energetics and NEB results provide a quantitative corroboration of hypothesis~(b): the blister creates a self-reinforcing mechanism that attracts and immobilizes surface molecules, preventing access to new intercalation sites and quenching further interlayer penetration. This does not rule out a residual role for the clustering pathway tested under hypothesis~(a) during the earliest stages of deposition, but it establishes the blister-induced kinetic trap as the dominant mechanism governing sorption behavior once intercalation has begun. The electronic origin of this dual role is examined in the following section.
 
\subsection{Electronic Structure and Charge Redistribution}
\label{sec:CDD}

\begin{figure*}[!t]
\centering
\includegraphics[width=\textwidth]{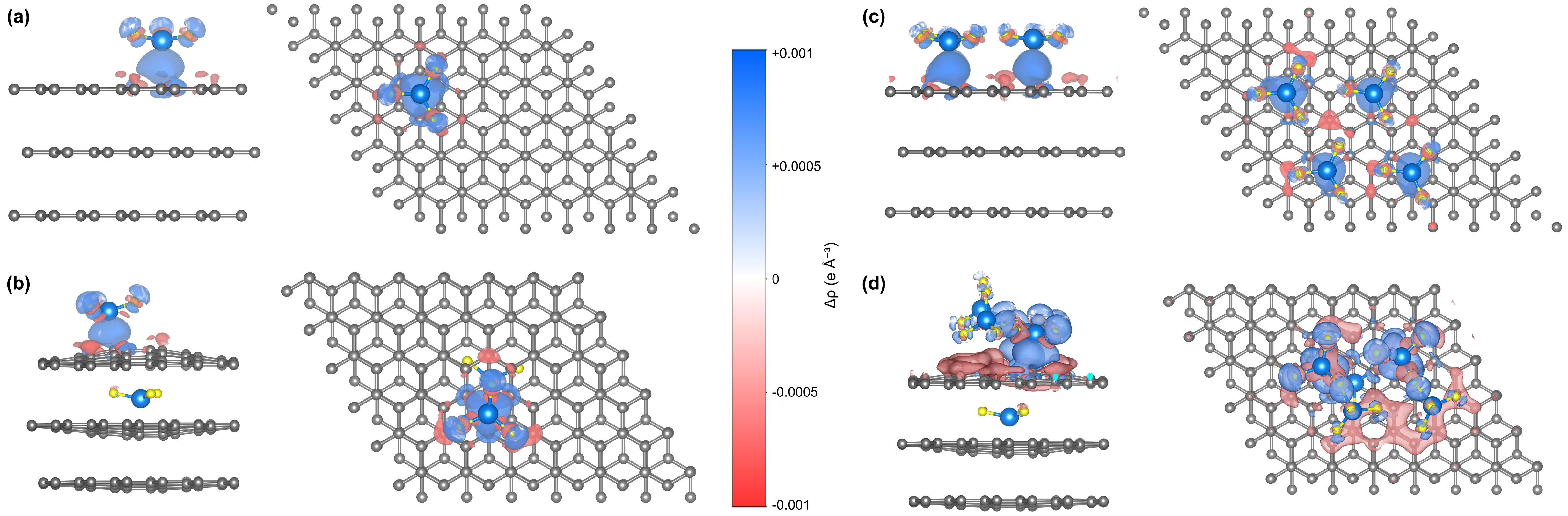}
\caption{Charge density difference $\Delta\rho$ (Eq.~\ref{eq:CDD}) for AlF$_3$ adsorption on (a)~pristine graphite $n=1$, (b)~blistered graphite $n=1$, (c)~pristine graphite $n=4$, and (d)~blistered graphite $n=4$. Each panel shows a side view (left) and top view (right). Blue isosurfaces ($\Delta\rho = +0.001$~e\,\AA$^{-3}$) indicate charge accumulation; red isosurfaces ($\Delta\rho = -0.001$~e\,\AA$^{-3}$) indicate charge depletion. Isosurface opacity: 180/255.}
\label{fig:CDD}
\end{figure*}

To identify the electronic origin of the cooperative adsorption enhancement and diffusion barrier increase discussed in Section~\ref{sec:diffusion}, we analyzed the charge density difference (CDD) and Mulliken population charges for representative configurations on pristine and blistered graphite. The CDD is defined as:
 
\begin{equation}
  \Delta\rho = \rho_{\mathrm{total}} - \rho_{\mathrm{graphite}} - \rho_{\mathrm{AlF_3}},
  \label{eq:CDD}
\end{equation}
 
where each component is evaluated at the geometry of the full system. Positive values (blue isosurfaces, $\Delta\rho = +0.001$~e\,\AA$^{-3}$) indicate charge accumulation; negative values (red isosurfaces, $\Delta\rho = -0.001$~e\,\AA$^{-3}$) indicate charge depletion. The resulting CDD maps are shown in Figure~\ref{fig:CDD} and the Mulliken charge analysis is summarized in Table~\ref{tab:mulliken}.
 
On the pristine surface, the interaction is highly localized: a single molecule perturbs only 13 carbon atoms and transfers 0.280~e to the graphite ($q_{\mathrm{Al}} = +1.232$~e, Table~\ref{tab:mulliken}). In the side view of Figure~\ref{fig:CDD}a, charge accumulation (blue) is concentrated directly beneath the Al atom, with charge depletion (red) confined to a narrow region around the fluorine atoms; the top view shows this perturbation restricted to the carbon atoms immediately below the molecule, with no discernible signal beyond the first coordination shell. As coverage increases to $n=4$ (Figure~\ref{fig:CDD}c), the accumulation regions remain isolated from one another in both side and top views---each molecule retains its own localized blue lobe with no overlap between neighbouring sites---and the charge transfer per molecule decreases to 0.142~e while the number of perturbed carbons rises to 56, reflecting competition among molecules for the same graphite donation sites.
 
In contrast, the intercalated AlF$_3$ molecule alone transfers 0.552~e and perturbs 61 carbon atoms, establishing a pre-existing charge redistribution that fundamentally alters the surface electronic landscape. On the blistered surface with a single surface molecule (Figure~\ref{fig:CDD}b), the side view reveals a markedly more extended and intense accumulation region than in the pristine case, with an additional charge redistribution visible around the subsurface intercalant itself; the top view shows the perturbation spreading beyond the immediate adsorption footprint, consistent with the larger charge transfer (0.463~e per molecule) and the increase to 70 perturbed carbons. At $n=4$ on the blistered surface (Figure~\ref{fig:CDD}d), the side view shows the individual accumulation regions of the four surface molecules merging into a single, continuous blue zone that spans the blister, in stark contrast to the discrete, isolated lobes of the pristine $n=4$ case (Figure~\ref{fig:CDD}c); the top view confirms this merged network extends across 89 perturbed carbons, with the charge transfer per molecule remaining at 0.302~e---more than twice the pristine value at equivalent coverage. Notably, the charge on the intercalated Al atom remains essentially constant at $\sim$0.96~e across all surface coverages, confirming that the intercalated molecule acts as a stable electronic reservoir that modifies the graphite potential without being significantly perturbed by the adsorbates above it.
 
\begin{table}[ht]
\centering
\caption{Mulliken charge analysis for AlF$_3$ adsorption on pristine and blistered graphite. $q_{\mathrm{Al}}$: charge on the surface Al atom (average for multiple molecules); $\Delta q_{\mathrm{Al}}$: charge variation relative to isolated AlF$_3$ ($q_{\mathrm{Al}}^{\mathrm{free}} = +1.463$~e); CT/mol: charge transfer per AlF$_3$ unit; $N_C$: number of significantly perturbed carbon atoms, defined as those with $|\Delta q_{\mathrm{C}}| > 0.005$~e, where $\Delta q_{\mathrm{C}}$ is the Mulliken charge 
variation of each carbon relative to the neutral-atom reference. This threshold filters out minor numerical fluctuations while retaining atoms undergoing meaningful electronic redistribution. For dimeric species, the reference is isolated Al$_2$F$_6$ ($q_{\mathrm{Al,sup}}^{\mathrm{free}} = +1.367$~e, $q_{\mathrm{Al,bridge}}^{\mathrm{free}} = +1.508$~e).}
\label{tab:mulliken}
\resizebox{0.7\columnwidth}{!}{%
\begin{tabular}{llcccc}
\toprule
System & Configuration & $q_{\mathrm{Al}}$ (e) & $\Delta q_{\mathrm{Al}}$ (e) & CT/mol (e) & $N_C$ \\
\midrule
\multicolumn{6}{l}{\textit{Pristine graphite --- monomers}} \\
& $n=1$ & $+1.232$ & $-0.231$ & $0.280$ & 13 \\
& $n=2$ & $+1.287$ & $-0.176$ & $0.198$ & 24 \\
& $n=3$ & $+1.301$ & $-0.162$ & $0.169$ & 41 \\
& $n=4$ & $+1.322$ & $-0.141$ & $0.142$ & 56 \\
& $n=5$ & $+1.315$ & $-0.148$ & $0.138$ & 60 \\
& $n=6$ & $+1.332$ & $-0.131$ & $0.114$ & 67 \\
\midrule
\multicolumn{6}{l}{\textit{Pristine graphite --- dimers}} \\
& 1 dimer, Al sup.   & $+1.344$ & $-0.023$ & $0.259$ & 21 \\
& 1 dimer, Al bridge & $+1.117$ & $-0.391$ & $0.259$ & 21 \\
& 2 dimers, Al sup.  & $+1.351$ & $-0.016$ & $0.221$ & 59 \\
& 2 dimers, Al bridge& $+1.143$ & $-0.365$ & $0.221$ & 59 \\
\midrule
\multicolumn{6}{l}{\textit{Blistered graphite}} \\
& intercalated only & $+0.976$ & $-0.487$ & $0.552$ & 61 \\
& $n=1$ surface     & $+1.176$ & $-0.287$ & $0.463$ & 70 \\
& $n=2$ surface     & $+1.338$ & $-0.125$ & $0.370$ & 74 \\
& $n=3$ surface     & $+1.295$ & $-0.168$ & $0.326$ & 85 \\
& $n=4$ surface     & $+1.243$ & $-0.220$ & $0.302$ & 89 \\
\bottomrule
\end{tabular}}
\end{table}
 
The Mulliken analysis provides a direct electronic basis for the two central findings of this work. Although Mulliken charges are basis-set dependent, only relative trends are discussed here. First, the cooperative adsorption strengthening on the blistered surface (Table~\ref{tab:adsorption_comparison}) is quantitatively reflected in the blister-amplified charge transfer: 65\% larger at $n=1$ and 113\% larger at $n=4$ relative to pristine graphite. Second, the greater diffusion barriers on the blistered surface (Table~\ref{tab:neb_sb_free})---equivalent to deeper potential wells at the adsorption sites the NEB pathways connect---coincide with a more extended charge perturbation (70--89 perturbed carbons vs.\ 13--67 on pristine graphite). While the two calculations were performed independently, their spatial correspondence suggests that the same charge redistribution responsible for the cooperative adsorption strengthening also underlies the kinetic trapping observed via NEB. 

This contrast is further corroborated by the projected density of states analysis (Section~S3, Supporting Information): on pristine graphite the $\pi$/$\pi^*$ bands exhibit a rigid shift, consistent with an intense perturbation of a few carbon atoms, whereas on the blistered surface they broaden without a pronounced shift, reflecting a redistribution over more carbons.
 
\begin{figure*}[!htb]
\centering
\includegraphics[width=0.7\textwidth]{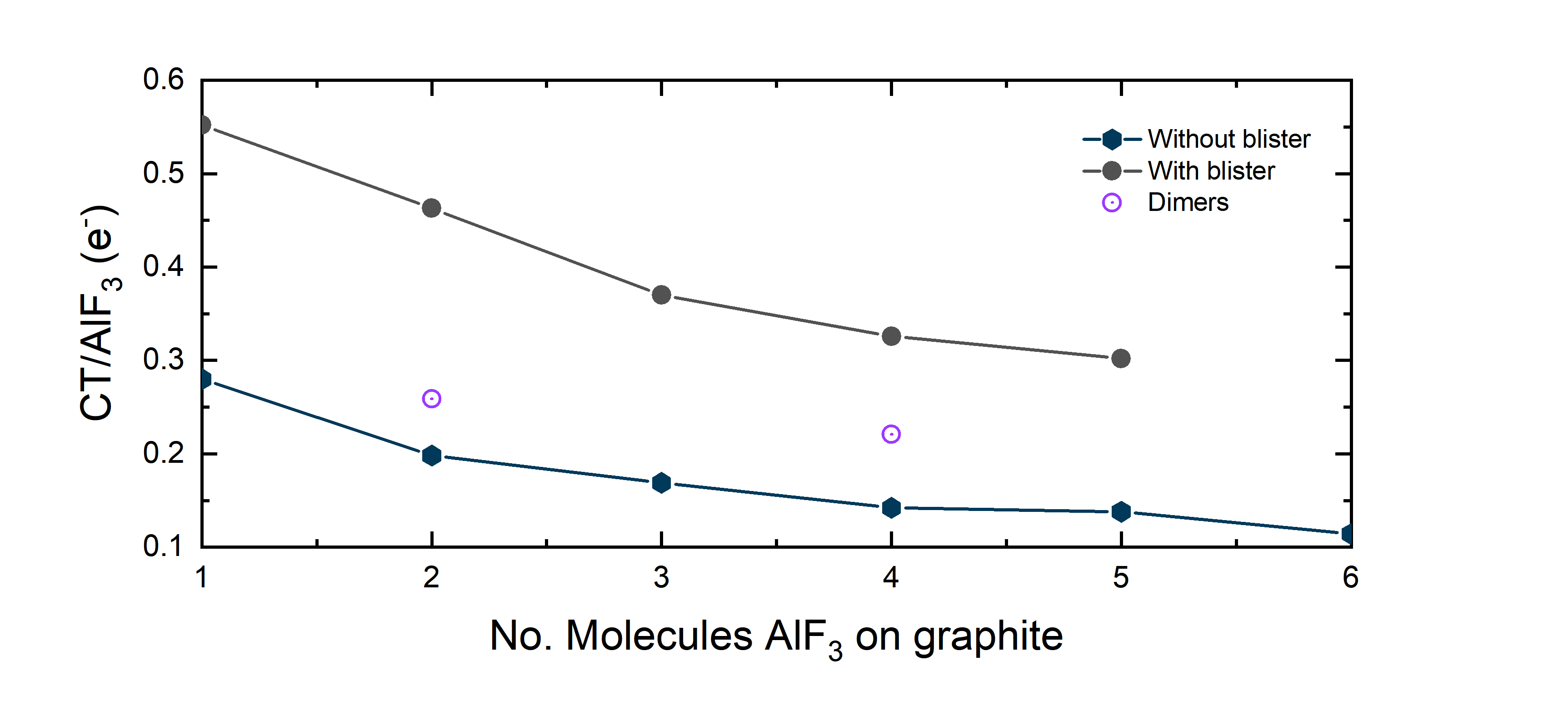}
\caption{Charge transfer per AlF$_3$ unit (CT/AlF$_3$) as a function of molecular coverage for pristine graphite (blue pentagons, "without blister") and blistered graphite (grey circles, "with blister"). Lilac-centered circles indicate dimeric configurations on the pristine surface. On the pristine surface, CT/AlF$_3$ decreases monotonically with coverage, reflecting competition among molecules for the same graphite donation sites. On the blistered surface, CT/AlF$_3$ remains systematically larger by a factor of $\sim$1.7--2.1$\times$ across all coverages, confirming the cooperative electronic enhancement induced by the subsurface intercalant.}
\label{fig:CT_coverage}
\end{figure*}

The full coverage dependence of this effect is summarized in Figure~\ref{fig:CT_coverage}. On the pristine surface (blue pentagons), CT/AlF$_3$ decreases monotonically from 0.280~e at $n=1$ to 0.114~e at $n=6$, consistent with the progressive dilution of the graphite donation capacity among an increasing number of adsorption sites discussed above. On the blistered surface (grey circles), CT/AlF$_3$ starts substantially higher and follows the same qualitative decreasing trend with coverage, yet remains systematically above the pristine curve across the entire coverage range, by a factor of $\sim$1.7--2.1$\times$. This persistent offset between the two curves---rather than a convergence at high coverage---is the clearest single indication that the intercalated molecule sustains a distinct, extended electronic environment that does not saturate on the same footing as the localized pristine sites. Dimeric configurations on the pristine surface (lilac-centered circles) fall above the pristine monomer curve at equivalent coverage, consistent with the internal charge mediation by the bridging Al atom discussed above, but remain below the blistered-surface values, confirming that the blister---not molecular clustering---is the dominant electronic driver of the cooperative enhancement.
 
These three independent lines of evidence---energetic, kinetic, and electronic---converge on a consistent atomistic picture that corroborates hypothesis~(b) as the unifying mechanism, while remaining compatible with a subordinate, kinetically limited role for the clustering pathway examined under hypothesis~(a). The electronic analysis presented here identifies the microscopic origin of the diffusion barriers discussed in Section~\ref{sec:diffusion}: it is the extended, coverage-independent charge reservoir established by the intercalant that deepens the surface potential landscape and ultimately drives the kinetic trapping responsible for the crossover from the fast to the slow exponential component in the biexponential AES fits (Figure~\ref{fig_1}).

\section{Conclusions}
Motivated by the experimentally observed flux-dependent biexponential sorption kinetics revealed by AES, the results presented here show that surface adsorption and subsurface intercalation of AlF$_3$ in graphite are not independent, competing processes but are coupled through a self-reinforcing structural and electronic feedback loop. Once a subsurface molecule deforms the graphite lattice into a blister, the modified surface becomes simultaneously more attractive for incoming adsorbates and more resistive to their lateral diffusion---a dual role that naturally explains the self-limiting biexponential kinetics observed experimentally in both LMF and HMF deposition regimes~\cite{Betancourt2026}.
 
Of the two hypotheses tested, molecular clustering (hypothesis~a) was not supported as an independent, thermodynamically driven pathway on the pristine surface: neither spontaneous monomer accumulation nor preformed dimers stabilized adsorption relative to isolated monomers. Clustering does occur spontaneously, but only once a blister is already present, consistent with it being a downstream consequence of the cooperative electrostatic environment established by the intercalant rather than an independent driver of enhanced coverage. Intercalation-induced surface conditioning (hypothesis~b) instead provides the microscopic mechanism underlying the flux-dependent balance between the fast and slow exponential components identified experimentally, operating across both flux regimes with a weight that shifts according to the relative rates of intercalation and surface accumulation.
 
This blister-induced trapping mechanism, as characterized here for the AlF$_3$--graphite system, operates through three cooperative channels: geometric confinement that promotes molecular clustering once intercalation has occurred, cooperative charge redistribution that deepens surface binding with increasing coverage, and kinetic arrest that prevents adsorbed species from reaching new intercalation sites. The proposed mechanism further predicts preferential AlF$_3$ accumulation around intercalation-induced blisters, providing a spatial signature that could be explored in future local-probe microscopy studies. Together, these effects transition the system from a quasi-barrierless to a thermally activated diffusion regime, with the contrast between pristine and blistered surfaces growing most pronounced at low temperatures.
 
These findings identify subsurface intercalation-induced deformation as a structural, kinetic and electronic mechanism capable of self-limiting further intercalation in graphite, complementing the sorption kinetics reported experimentally for this system. Given the established role of AlF$_3$ as an interfacial modifier in carbon-based battery electrodes, the blister-induced trapping mechanism characterized here may be a relevant consideration for the design and processing of AlF$_3$-modified carbon electrodes, where deposition flux and substrate defect density emerge as tunable parameters for controlling the balance between surface adsorption and interlayer intercalation.

\section*{CRediT authorship contribution statement}

\textbf{H.S. Betancourt Infante:} Conceptualization, Methodology, Investigation, Formal analysis, Data curation, Visualization, Validation, 
Writing -- original draft.
\textbf{F.J. Bonetto:} Conceptualization, Formal analysis, Supervision, Project administration, Funding acquisition, Validation, Writing -- review \& editing.
\textbf{G.D. Ruano:} Conceptualization, Methodology, Investigation, Formal analysis, Supervision, Funding acquisition, Validation, Writing -- original draft.
\textbf{S.J. Rodr\'iguez-Sotelo:} Methodology, Investigation, Formal analysis, Visualization, Validation, Writing -- original draft.

\section*{Declaration of competing interest}

The authors declare that they have no known competing financial interests or personal relationships that could have appeared to influence the work reported in this paper.
\section*{Acknowledgements}

This work was supported by CONICET through grants PIP-2021-101517, PIP-112\-202\-001\-002\-57CO, and PIP-112\-202\-001\-003\-84CO; by Universidad Nacional del Litoral (UNL) through CAI+D grants No.~85520240100061LI and No.~2020-501\-901\-001\-64LI; by ASaCTeI through grant PEICA-2023-039; and by ANPCyT through PICT grants 2019-04545, 2019-03493, 2020-01193, and 2021-I-INVI-00863.

H.S.B.\ acknowledges a doctoral fellowship from CONICET. H.S.B.\ and G.R.\ acknowledge secondment grants from the European Union's Horizon~2020 research and innovation programme under the Marie Sk\l{}odowska-Curie grant agreement No.~101007825 (ULTIMATE-I, H2020-MSCA-RISE-2020), carried out at the Universidad de Zaragoza, Spain. F.B.\ acknowledges financial support from the European Union under the REFRESH -- Research Excellence For Region Sustainability and High-tech Industries project No.~CZ.10.03.01/00/22\_003/0000048 via the Operational Programme Just Transition, and from ENREGAT supported by M\v{S}MT, project No.~LM2023056.

Computational resources were provided by the Pirayu cluster, acquired with funds from the Agencia Santafesina de Ciencia, Tecnolog\'ia e Innovaci\'on (ASACTEI), Government of the Province of Santa Fe, through Project AC-00010-18, Resolution No.~117/14. This equipment is part of the National High-Performance Computing System of the Ministry of Science and Technology of Argentina. The authors gratefully acknowledge the C3ADMIN support team of the Pirayu cluster---J.\ Gonz\'alez, F.\ Lell, and F.\ Lezcano, Centro de Investigaci\'on de M\'etodos Computacionales (CIMEC)--CONICET---for their technical assistance throughout this work, with particular thanks to F.\ Lezcano.

\bibliographystyle{unsrt}  
\bibliography{cas-refs}

\end{document}